\documentclass[superscriptaddress,reprint,longbibliography]{revtex4-1}
\usepackage{amssymb,bbold,dsfont}
\usepackage{amsmath}
\usepackage{color}
\usepackage{graphicx}
\usepackage{marvosym}
\usepackage{ulem}
\usepackage{notoccite}
\usepackage{comment}
\usepackage{pifont}
\usepackage{bbm}
\usepackage{graphicx,tabularx}
\usepackage{dcolumn}
\usepackage{bm}
\usepackage{xcolor}
\usepackage{marvosym}
\usepackage[breaklinks=true, colorlinks,citecolor=blue,linkcolor=blue,urlcolor=blue]{hyperref}
\usepackage{titlesec}
\newcommand{\be}{\begin{equation}}
\newcommand{\ee}{\end{equation}}
\newcommand{\nn}{\nonumber\\}

\renewcommand{\bm}[1]{{\bf #1}}
\def \bea{\begin{eqnarray}}
\def \eea{\end{eqnarray}}
\def \nn{\nonumber}
\def \moire{moir\'e }

\begin{document}

\title{Large quantum nonreciprocity in plasmons dragged by drifting electrons}
    \author{Debasis Dutta}
    \email{ddebasis@iitk.ac.in}
    \affiliation{Department of Physics, Indian Institute of Technology 
    Kanpur, Kanpur-208016, India}
   \author{Amit Agarwal}
   \email{amitag@iitk.ac.in}
   \affiliation{Department of Physics, Indian Institute of Technology Kanpur, Kanpur-208016, India}
\date{\today}
\begin{abstract}

Collective plasmon modes, riding on top of drifting electrons, acquire a fascinating nonreciprocal dispersion characterized by $\omega_p(\bm{q}) \neq \omega_p(-\bm{q})$. The classical plasmonic Doppler shift arises from the polarization of the Fermi surface due to the applied DC bias voltage. {Here, we predict an additional quantum contribution to the plasmonic Doppler shift originating from the quantum metric of the Bloch wavefunction. We systematically compare the classical and quantum corrections to the Doppler shifts by investigating the drift-induced nonreciprocal plasmon dispersion in graphene and in twisted bilayer graphene. 
We show that the quantum plasmonic Doppler shift dominates in \moire systems at large wavevectors, yielding plasmonic nonreciprocity up to 20\% in twisted bilayer graphene. Our findings highlight the significance of the quantum corrections to plasmonic Doppler shift in \moire systems and motivate the design of innovative nonreciprocal photonic devices with potential technological implications.}
\end{abstract}
\maketitle


\section{Introduction}

Light propagates symmetrically in opposite directions in conventional optical systems. This is a consequence of the time-reversal invariance of Maxwell's equations or Lorentz's reciprocity principle~\cite{Caloz2018, RJPotton_2004, Boriskina2022, Tokura2018N}. Breaking reciprocity for asymmetric light propagation is conventionally done by magneto-optical approaches, which require large magnetic fields and limit the efficiency for nanoscale devices and on-chip integration~\cite{Bi2011O}. To remedy this, nonreciprocal plasmonics in atomically thin two-dimensional (2D) materials, such as graphene, present opportunities for direction-dependent light propagation at the nanoscale. This is crucial for enabling compact devices in classical and quantum information processing, nonreciprocal devices for Faraday rotation, isolation, one-way waveguiding, and nonreciprocal cavities~\cite{Boyd2017B, Monticone2020, HassaniGangaraj2022}. These make them a valuable addition to the nanophotonics toolbox~\cite{DNBasov2016, FrankHKoppen2021B, carlo2021}~.
%
\begin{figure}[t!]
	\includegraphics[width=\linewidth]{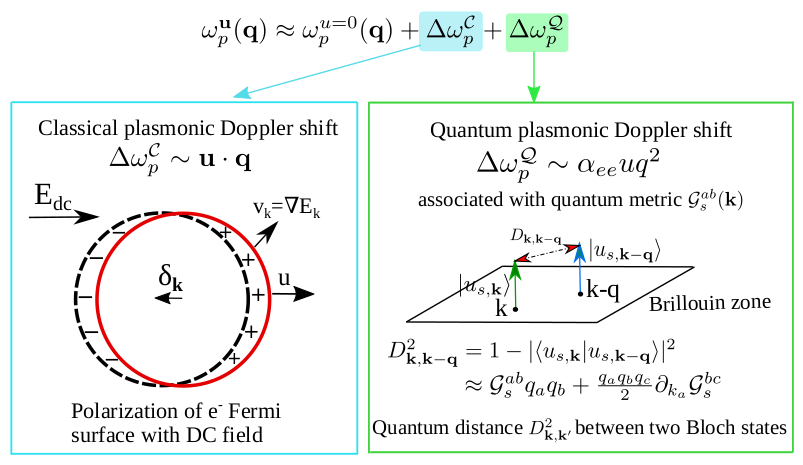}
	\caption{The drift-induced nonreciprocal plasmon dispersion has both classical and quantum contributions. 
        The classical frequency shift $\Delta \omega_p^{\cal C}$  arises from the polarization of the Fermi surface induced by the DC electric field. The solid and dashed line represents the Fermi surface in the absence and presence of drift flow ($u$) of carriers. 
        The quantum plasmonic Doppler shift, $\Delta \omega_p^{\cal Q}$, arises from the quantum metric ${\cal G}^{ab}_s(\bm{k})$ of the electron wavefunction in the presence of drifting charge carriers. 
        The quantum metric is linked with the notion of quantum distance, $D^{2}_{\bm{k},\bm{k}-\bm{q}}$ between two Bloch states at different momentum, $\bm{k}$ and $\bm{k}$$-$$\bm{q}$, respectively. 
        }
 \label{Fig1}
\end{figure}

Nonreciprocity in bulk plasmon dispersion can intrinsically arise in noncentrosymmetric magnetic materials. This is induced either by the dipolar distribution of the quantum metric or by the `chiral Berry' plasmons at the boundary of magnetic materials~\cite{Song2015, Justin_song2022, Huang2022, Dutta2023I}. A more promising and controllable route for extrinsic breaking of Lorentz reciprocity is biasing the plasmonic material with a direct current. This induces nonreciprocal plasmons with a dispersion that differs for plasmon propagating along or opposite to the direction of the drifting carriers. This approach is minimally invasive for to on-chip architectures, and it is known as the plasmonic Fizeau drag or Doppler effect~\cite{DNBasov2021, Zhao2021, Levitov2015Q, Lewandowski2020, HassaniGangaraj2022, MorgadoSilveirinha2022}.
Drift-induced nonreciprocal plasmons in single-layer graphene (SLG) have been recently predicted and demonstrated \cite{Levitov2015Q, Marco_2016, Stauber2015, DNBasov2021, Morgado2020N, Silveirinha2022D}. Using near-field imaging techniques, the plasmonic Doppler shift was measured for SLG~\cite{DNBasov2021, Zhao2021} with a wavelength shift, $\delta \lambda_p/{\lambda_p}\approx{2}\%$ for electron drift velocity, $u/v_F\approx 17\%$ (given $v_F=0.86\times{10}^6$ m/s)~\cite{DNBasov2021}. These studies are primarily focussed on the classical plasmonic Doppler shift, which primarily arises from the displacement of the Fermi surface due to drift flow (see Fig.~\ref{Fig1})
~\cite{Levitov2015Q, Marco_2016, Stauber2015, DNBasov2021, Morgado2020N, Silveirinha2022D}.
{Additionally, the possibility of a quantum Doppler shift has been recently proposed in \moire system owing to the band hybridization ~\cite{Lewandowski2020}.}

{Motivated by these studies, we predict an exciting quantum Doppler shift-induced plasmonic nonreciprocity, which originates from the quantum metric - a band geometric property of the electron wavefunctions (see Fig.~\ref{Fig1})}. 
Our investigation reveals that the classical correction varies linearly with the wavevector, while the quantum correction varies quadratically with the wavevector, as depicted in Fig.~\ref{Fig1}. We show that as flat band \moire systems have a large effective interaction strength and undamped plasmons at large wavevectors, twisted bilayer graphene (TBG) can support a large plasmonic nonreciprocity ($\sim$20\%) driven by quantum correction. As a consequence, \moire materials in general and TBG in particular offers highly tunable platforms for observing and designing devices based on nonreciprocal light propagation ~\cite{Provost1980, Bghosh2023Q, PhysRevLett.127.246403, Agarwal2022A, Levitov2019}.

\section{Nonreciprocal plasmons with DC bias}

The optical and plasmonic properties of a quantum system can be described by the dynamical density-density response function in the linear response of the applied electric field~\cite{giuliani2005quantum, Pines1962, Dutta_C_2022}. Plasmons are calculated from the 
zeros of the real part of the dielectric function. 
Within random-phase approximation (RPA), the dynamical dielectric function is calculated as~\cite{giuliani2005quantum, PhysRevB.90.155409, PhysRevB.91.245407}
\bea
\varepsilon(\bm{q},\omega)= 1 - V_{q}\Pi(\bm{q},\omega)~.
\label{epsilon_RPA}
\eea
Here, $V_{q}=2\pi e^2/(\kappa |\bm{q}|)$ denotes the 2D Fourier transform of the Coulomb potential, $\kappa$ denotes an effective background dielectric constant, and  $\Pi(\bm{q},\omega)$ represents dynamical density-density response function~\cite{giuliani2005quantum, RashiSachdeva2015, Dutta_C_2022}. 

We will treat the effect of externally applied DC current in a non-perturbative fashion by capturing its impact on the Fermi-Dirac distribution function~\cite{Marco_2016, Lewandowski2020, Stauber2015}. 
{In the hydrodynamic limit}, we can model the drifting carrier distribution function as~\cite{MacDonald2009, Gantmakher1988, DNBasov2021}
\be
\tilde{f}_{s,\bm{k}}=\left\{ {\rm exp}\left[\frac{E_{s,\bm{k}}-\bm{u}\cdot{\bm k}-\mu }{k_BT} \right]+1  \right\}^{-1}~,
\label{f_tilde_u}
\ee
which nullifies the energy and momentum conserving electron-electron collision integral~\cite{MacDonald2009}. Here, $\bm{u}$ denotes the drift velocity of the quasiparticles, $T$ denotes temperature, $\mu$ denotes the chemical potential, and $E_{s,\bm{k}}$ denotes Bloch band energy at crystal momentum $\bm{{k}}$ with band index $s$ of the system. This drifting carrier distribution function induces a shift of the Fermi surface by momentum $\delta_{\bm{k}}=-m_{\rm eff}~\bm{u}$, in the small $u =|{\bm u}|$ limit~\cite{Lewandowski2020} {(see Sec.~S1 of the {Supplemental Material} (SM)~\footnote{\href{https://www.dropbox.com/scl/fi/ypgttk1ocoxpm43qmrs0p/SM_final.pdf?rlkey=fandngfoygkd0fdyf5ah68qb5&st=90ayzzew&dl=0}{The Supplemental material} discusses, i) Eq. 2 in different regimes,  ii) The density-density response function in small $q$ limit, iii) Calculation of nonreciprocal plasmon dispersion, iv) Hydrodynamic theory of quantum plasmonic Doppler shift, v) Nonreciprocal plasmons in 2DEG, vi) Nonreciprocal plasmon dispersion in graphene, vii) Continuum model Hamiltonian for twisted bilayer graphene, viii) Effective Fermi velocity in twisted bilayer graphene} for details)}. Here, $m_{\rm eff}$ denotes the effective mass of the quasiparticles.

Using this approach, we can express the DC current-driven non-interacting density-density response function for a  2D system  as~\cite{giuliani2005quantum, Marco_2016, PhysRevB.83.115135}
\be
\Pi(\bm{q},\omega)=g_s\sum_{s,s^{\prime}}\int \frac{d^2\bm{k}}{(2\pi)^2}\frac{\left(\tilde{f}_{s,\bm{k}+\bm{q}} -\tilde{f}_{s^{\prime},\bm{k}} \right) F^{ss^{\prime}}_{\bm{k}+\bm{q},\bm{k}} }{E_{s,\bm{k}+\bm{q}}- E_{s^{\prime},\bm{k}}-\omega -i\eta}~.
\label{Pi_Eq1}
\ee
Here, $\tilde{f}_{s,\bm{k}}$ is the modified Fermi-Dirac distribution function at momentum $\bm{{k}}$ with band energies $E_{s,\bm{k}}$, $g_s$ denotes the total degeneracy factor, and $\eta$ is the broadening parameter. Here, the important quantity is the band coherence factor $F^{ss^{\prime}}_{\bm{k}+\bm{q},\bm{k}}=|\langle u_{s,\bm{k}+\bm{q}}|u_{{s}^{\prime},\bm{k}}\rangle|^2$, which describes the overlap between two energy eigenstates at momentum $\bm{k}$ and $\bm{k}+\bm{q}$. We set $\hbar=1$ throughout our calculations and explicitly mention it when needed. 

We first investigate the long-wavelength limit ($q\ll k_F$, where $k_F$ denotes the Fermi wavevector) of the intraband plasmon dispersion in the presence of drift flow. For that, we expand the intraband band-overlap factor, $F^{ss}_{\bm{k},\bm{k}+\bm{q}}$ up to ${\mathcal O}(q^3)$~\cite{Justin_song2022, ROSSI2021Q}, and obtain 
\be
F^{ss}_{\bm{k\pm q},\bm{k}}\approx 1- q_aq_b{\cal G}^{ab}_{s} \mp \frac{q_aq_bq_c}{2}\partial_{k_a}{\cal G}^{bc}_{s}~.
\label{Eq_F}
\ee
Here, ${\cal G}^{ab}_{s}(\bm{k})=\left[ {\rm Re}\langle \partial_{k_a}u_{s,\bm{k}}|\partial_{k_b} u_{s,\bm{k}}\rangle -\xi^{a}\xi^{b} \right]$ represents intraband quantum metric (or the 
Fubini-Study metric), with $\xi^a=i\langle u_{s,\bm{k}}|\partial_{k_a}u_{s,\bm{k}}\rangle$ being the single band Berry connection~\cite{Provost1980, Justin_song2022, Xiao_PRL2019, Resta2011}, and $a$, $b$, $c$ denote cartesian directions. The quantum metric measures the distance between two infinitesimally close Bloch states in Hilbert space as~\cite{Resta2011, PhysRevResearch.5.L012015}
$D^2_{\bm{k},\bm{k}+d\bm{k}}=1-|\langle u_{s,\bm{k}}|u_{s,\bm{k}+d\bm{k}}\rangle|^2\simeq{\cal G}^{ab}_{s}(\bm{k})d\bm{k}^ad\bm{k}^b$.

We work in the dynamical long-wavelength limit, $qv_F< \omega \ll \mu$ ($v_F$ denotes Fermi-velocity) to probe long-wavelength plasmons. In this limit,  we expand the real part of the density-density response function in different orders of $1/\omega$. The calculation details are discussed in Sec.~\textcolor{blue}{S2} of SM. We obtain,  
\be
{\rm Re}\left[ \Pi_{\rm intra}(\bm{q},\omega)\right] = q_aq_bq_c\frac{Q^{\bm{u}}_{abc}}{\omega} +q_aq_b\frac{{\cal D}^{\bm{u}}_{ab}}{\omega^2} + q_aq_bq_c\frac{C^{\bm{u}}_{abc}}{\omega^3}+\cdots
\label{Pi_Eq5}
\ee
We have used the $\bm{u}$-superscript to denote their drift current dependence. The expansion coefficients of Eq.~\ref{Pi_Eq5} are specified by, 
\bea
Q^{\bm{u}}_{abc} &=& -g_s\sum_{s,\bm{k}}\tilde{f}_{s,\bm{k}}\partial_{k_a}{\cal G}^{bc}_{s}(\bm{k})~,
\label{Eq_A1a}
\\
{\cal D}^{\bm{u}}_{ab} &=&  g_s\sum_{s,\bm{k}} \tilde{f}_{s,\bm{k}}\left( \frac{\partial^2 E_{s,\bm{k}}}{\partial k_a \partial k_b} \right)~,
\label{Eq_A2a}
\\
C^{\bm{u}}_{abc} &=& 2g_s\sum_{s,\bm{k}}\tilde{f}_{s,\bm{k}}\left(v_{s,\bm{k}}^{a} \frac{\partial^2 E_{s,\bm{k}}}{\partial k_b \partial k_c} \right)~.
\label{Eq_A3a}
\eea
Here, $v^{a}_{s,\bm{k}}=\partial E_{s,\bm{k}}/\partial k_a$ represents the band velocity component. 
 ${Q}^{\bm{u}}_{abc}$ is the drift-induced quantum-metric dipole~\cite{Justin_song2022, Xiao_PRL2019, Bghosh2023Q, Hughes2019}. It is analogous to the  Berry curvature dipole~\cite{Sodemann2015Q}, with the Berry curvature~\cite{Sodemann2015Q} substituted by the quantum metric. In Eq.~\eqref{Eq_A2a}, ${\cal D}^{\bm{u}}_{ab}$ is the drift-renormalized Drude-weight. In Eq.~\eqref{Eq_A3a}, ${\cal C}^{\bm{u}}_{abc}$ arises from the polarization of the Fermi-surface due to the momentum shift ($\delta_{\bm{k}}$) induced by the DC bias. 

 We find that the odd $1/\omega$ expansion coefficients in Eq.~\eqref{Eq_A1a} and Eq.~\eqref{Eq_A3a} are crucial for supporting nonreciprocal plasmons in the systems with $\Pi(\bm{q},\omega)\ne \Pi(-\bm{q},\omega)$.
 In the absence of drift, ${Q}^{\bm{u}}_{abc}$ and ${\cal C}^{\bm{u}}_{abc}$ will be identically zero unless the system intrinsically breaks the inversion and time-reversal symmetry, simultaneously~\cite{Justin_song2022, Dutta2023I}. 
In that case, ${\cal Q}^{\bm{u}=0}_{abc}$, and ${\cal C}^{\bm{u}=0}_{abc}$ give rise to intrinsic nonreciprocal plasmon modes in noncentrosymmetric magnetic systems~\cite{Justin_song2022,Dutta2023I}. However, the unidirectional flow of drifting electrons breaks both the time-reversal and inversion symmetry, and it leads to an asymmetric Fermi-distribution function, $\tilde{f}_{s,\bm{k}}\ne \tilde{f}_{s,-\bm{k}}$. Therefore, the $\bm{k}$-integration of Eq.~\eqref{Eq_A1a}, and Eq.~\eqref{Eq_A3a} are non-zero, yielding finite values of ${\cal Q}^{\bm{u}}_{abc}$ and ${\cal C}^{\bm{u}}_{abc}$ in presence of finite $u$ in all quantum systems.

We obtain the plasmon frequency by solving for the roots of Eq.~\eqref{epsilon_RPA} on the real axis~\cite{giuliani2005quantum}, assuming the Landau damping to be relatively small. 
Retaining terms up to $\omega^3$ in Eq.~\eqref{Pi_Eq5}, we can approximate the long wavelength nonreciprocal plasmon dispersion. We calculate the drift-induced nonreciprocal plasmon dispersion for small $q$, up to linear order in $u$, to be (see Sec.~\textcolor{blue}{S3} of SM for details)
\be
\omega_{p}^{\bm{u}}(\bm{q})\approx \sqrt{q^2V_q{\cal D}^{\bm{u}}} + \Delta\omega_p^{\cal C} + \Delta\omega_p^{\cal Q}~.
\label{intra_non_reciprocity}
\ee
Here, $\Delta\omega_p^{\cal C}=q{\cal C}^{\bm{u}}/2{\cal D}^{\bm{u}}$ captures the classical plasmon Doppler shift, and $\Delta\omega_p^{\cal Q}=q^3V_q{\cal Q}^{\bm{u}}/2$ is the quantum plasmon Doppler shift. 

Equation~\eqref{intra_non_reciprocity} captures the long-wavelength limit of nonreciprocal plasmon dispersion for general quantum systems with a DC current (retaining terms up to linear order in $\bm{u}$). Here, the first term captures the reciprocal plasmon dispersion with a drift velocity modified Drude weight, ${\cal D}^0\to {\cal D}^{\bm{u}}$. Both of the other terms capture the nonreciprocal dispersion, as the sign of ${\cal Q}^{\bm{u}}$, and ${\cal C}^{\bm{u}}$ depends on whether the plasmon propagates along or opposite to the drift flow, {\it i.e.}, $\bm{\hat{\bm{q}}}\cdot{\hat{\bm{u}}}=\pm{1}$. Of these two, the $\Delta\omega_p^{\cal C}$ term is a linear-in-$q$ correction while the $\Delta\omega_p^{\cal Q}$ captures a quadratic correction to the nonreciprocal plasmon dispersion (for unscreened Coulomb interactions). 
The frequency shift $\Delta\omega_p^{\cal C}$ arises from the drift velocity induced shift of the Fermi-surface, and it is typically referred to as classical correction or classical plasmonic Doppler shift~\cite{gao2019plasmonic}. In contrast, the second correction term, $\Delta\omega_p^{\cal Q}$ in Eq.~\eqref{intra_non_reciprocity} has a completely quantum origin associated with the quantum-metric dipole, ${\cal Q}^{\bm{u}}$. We term this correction as quantum plasmonic Doppler shift, as it arises from the nontrivial quantum geometry of the Bloch state~\cite{Di_XiaoRMP2010}. {This additional correction to the Doppler shift is one of the main findings of this manuscript. 
Interestingly, the quantum correction to the plasmonic Doppler shift can also be derived from a semiclassical hydrodynamic description. We present the semiclassical description of the plasmonic Doppler shift in Sec.~4 of SM}.
 
To quantify the classical and quantum nonreciprocity, we compute the percentage of nonreciprocity,
\bea
\frac{| \omega_p^{\bm{u}}(\bm{q})-\omega_p^{\bm{u}}(-\bm{q})|}{\omega_p^{0}(\bm{q})}&=& \underbrace{\sqrt{q}\frac{{\cal C}^{\bm{u}}}{{\cal D}^{\bm{u}}} \sqrt{\frac{\kappa}{2\pi e^2{\cal D}^0}}}_{\eta^{\cal C}} + \underbrace{q^{3/2}{\cal Q}^{\bm{u}}\sqrt{ \frac{2\pi e^2}{\kappa{\cal D}^0}}}_{\eta^{\cal Q}}~.\nn
\\
\eea
Here, $\eta^{\cal C}$ and $\eta^{\cal Q}$ denote the percentage of classical and quantum plasmonic nonreciprocity, respectively, and we have defined ${\cal D}^0 \equiv {\cal D}^{\bm u = 0}$. Having established the origin of drift-induced quantum nonreciprocity, we next explore the magnitude of these terms in two-dimensional electron gas (2DEG), graphene, and twisted bilayer graphene.

\section{Quantum nonreciprocity in graphene and 2DEG}
We calculate Eq.~\eqref{intra_non_reciprocity} for 2DEG with parabolic dispersion and graphene having linear dispersion to elucidate classical and quantum nonreciprocity. 
The wavefunction of a 2DEG is a single component object, and as a consequence, the band overlap term $F_{\bm{k},\bm{k}\pm\bm{q}}=1$. This can also be seen from the fact that the Berry curvature and quantum metric vanish in single-component systems.
Thus, a 2DEG can only support classical nonreciprocity with $\Delta \omega_p^{\cal C}=\bm{u}\cdot{\bm{q}}$, and the quantum contribution vanishes completely. 
We obtain the drift-induced plasmon dispersion for 2DEG from Eq.~\eqref{intra_non_reciprocity} to be, 
\bea
\omega_{\rm 2DEG}^{\bm{{u}}}(\bm{q})&\approx& \sqrt{\frac{2\pi n e^2}{\kappa m}q}+\bm{u}\cdot\bm{q}~.
\label{2DEG_omega_p}
\eea
Here, $n$ is the 2DEG carrier density and $m$ is the effective mass. The detailed derivations are shown in Sec.~\textcolor{blue}{S5} of SM~\cite{Note1}.
\begin{figure}[t!]
	\includegraphics[width=1.0\linewidth]{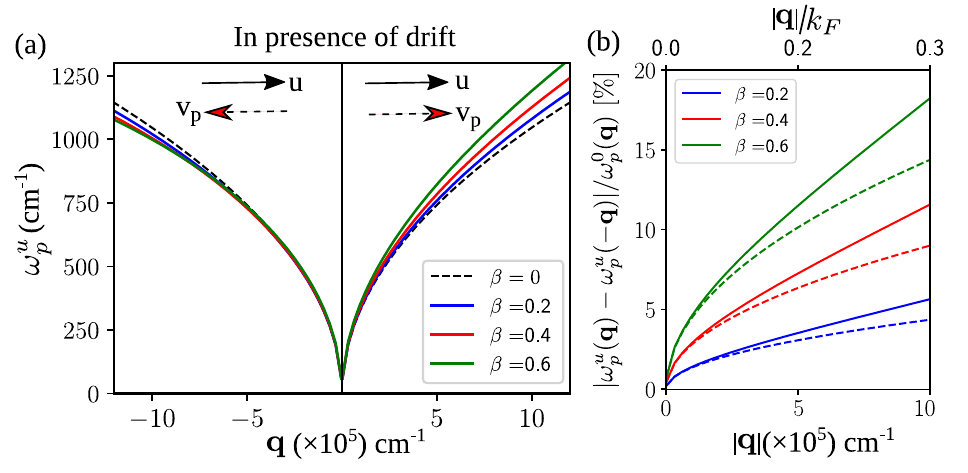}
	\caption{(a) Plasmon dispersion for current-carrying graphene under different carrier drift velocities, $\beta= u/v_F = 0.2$, $0.4$, and $0.6$, respectively. Here, v$_p=d\omega_p/dq$ indicates the group velocity of plasmons. We have used $\alpha_{\rm ee}=0.9$ for  air/SLG/SiO$_2$ interface~\cite{Grigorenko2012} and carrier density, $n=2.9\times 10^{12}$ cm$^{-2}$~\cite{DNBasov2021}. 
    (b) Total percentage of nonreciprocity, $|\omega^{u}_p(\bm{q})-\omega^{u}_p(-\bm{q})|/\omega_{p}^{0}(q)$ as a function of momentum ($\bm{q}$) for current-carrying graphene with different $\beta$. The dashed line represents the classical percentage of nonreciprocity ($\eta^{\cal C}$) as defined in Eq.~\eqref{Eq_classical}. {The horizontal $\rm{q}$-axis has been scaled in terms of the Fermi wavevector ($k_{\rm F}$) at the top of the panel (b).} For SLG, the percentage of nonreciprocity is mainly dominated by classical contribution $\Delta \omega_p^{\cal C}$. 
    }
    \label{Fig2_SLG}
\end{figure}

In contrast to 2DEG, the low energy electronic states for graphene is represented by 2D massless Dirac Hamiltonian, specified by ${\cal H}_{\bm{k}}= v_F{\sigma}\cdot{\bm{k}}$ with $\sigma=(\sigma_x,\sigma_y)$ being the vector of the Pauli matrices and $v_F$ is the Fermi velocity~\cite{Castroneto2009}. This Hamiltonian has two component spinor eigenstates, $|\bm{k},s\rangle=(1/\sqrt{2})\left( e^{-i\theta}~~s\right)^T$, with eigenvalues, $E_{s,\bm{k}}=sv_F|\bm{k}|$ for conduction ($s=1$) and valence ($s=-1$) band respectively, and $\theta=\tan^{-1}(k_y/k_x)$. 
To understand the role of quantum geometry, we calculate band resolved quantum metric, and it is given by ${\cal G}_{+}^{xx}(\bm{k})={\cal G}_{-}^{xx}(\bm{k})={\sin^2\theta}/(4k^2)$. The corresponding band overlap term can be evaluated using Eq.~\eqref{Eq_F}. For $\bm{q}=q\hat{\bm{x}}$, it is given by
\be
F^{++}_{\bm{k}\pm\bm{q},\bm{k}}\approx 1- \frac{q^2}{4k^2}\sin^2\theta \pm \cos\theta\sin^2\theta \frac{q^3}{2k^3}~.
\ee

To include the impact of the unidirectional drift flow $\bm{{u}}=u\hat{\bm{x}}$ in the Fermi-Dirac distribution, we model it for $T=0$ as $\tilde{f}_{+,\bm{k}}=\Theta[k_F(\theta)-k]$, where $k_F(\theta)=k_F/[1-\beta\cos(\theta-\phi_u)]$. Here, $\beta= u/v_F$, and $\phi_u$ is the angle between $\bm{u}$ and plasmon wavevector $\bm{q}$. 
With this modified Fermi-distribution function, we can calculate the quantum metric dipole to be ${\cal Q}^{\bm{u}}={\gamma}{g_su}/({16\pi|\mu|})$. 
Here, $g_s=4$ represent total spin and valley degeneracy, $\mu$ is the chemical potential, and $\gamma=\hat{\bm{q}}\cdot\hat{\bm{u}}$~. For example, $\gamma = +1~(-1)$ represents an up-stream (down-stream) plasmon propagation with respect to the drift of the charge carriers. We calculate the other expansion coefficients of $\Pi(\bm{{q}},\omega)$ in Eq.~\eqref{Pi_Eq5} to be, 
\be
{\cal D}^{\bm{u}}= \frac{g_s|\mu|}{4\pi}\frac{W(\beta)}{\beta}~,~~C^{\bm{u}}= \gamma\frac{g_sv_F|\mu|}{8\pi}\frac{W(\beta)^2}{\beta}~.
\ee
Here, we have defined $W(\beta)=2(1-\sqrt{1-\beta^2})/{\beta}$ and this relativistic factor becomes unity as $u \to 0$, or ${\rm lim}_{\beta\to 0}W(\beta)/\beta=1$~\cite{Marco_2016}. Combining these terms, 
we calculate the long wavelength drift-induced nonreciprocal plasmon dispersion for graphene to be 
(see Sec.~\textcolor{blue}{S6} of SM~\cite{Note1} for detailed calculations),
\bea
\omega_p^{\bm{u}}(\bm{q}) &\approx & \sqrt{\frac{2D_0 W(\beta)}{\kappa\beta}}\sqrt{q}\left[1+  \frac{12-16\alpha_{ee}^2}{16}\frac{q}{k_{\rm TF}} \right]~\nn
\\
&+& \gamma\frac{W(\beta)}{4\beta}uq +\gamma\frac{\alpha_{\rm ee}}{4k_F}uq^2~.
\label{Eq_12}
\eea
Here, $D_0=e^2\mu/{\hbar^2}$ is the non-interacting Drude-weight at $T=0$ for the 2D massless Dirac Hamiltonian~\cite{Grigorenko2012}, $\alpha_{ee}=e^2/(\kappa\hbar v_F)$ is the effective-fine structure constant, and $k_{\rm TF}=4\alpha_{\rm ee}k_F$ and $\kappa$ is the background dielectric constant. 
In Eq.~\eqref{Eq_12}, the last two nonreciprocal terms are calculated up to linear order in the drift velocity ~\cite{Marco_2016, DNBasov2021}.

In Eq.~\eqref{Eq_12}, the classical Doppler shift due to ${\cal C}^{u}$ is $\Delta \omega_{p}^{\cal C}= {\gamma}[{W(\beta)}/{4\beta}]uq$, which arises from the polarization of Fermi surface under DC field. Additionally, there is a quantum correction in Eq.~\eqref{Eq_12} originating from the quantum-metric dipole ${\cal Q}^{\bm{u}}$, and this quantum Doppler shift is given by $\Delta \omega_p^{\cal Q}= \gamma({\alpha_{\rm ee}}/{4k_F})uq^2$. This quantum correction goes as $q^2$ and varies with the carrier density as $n^{-1/2}$. 
We calculate the classical and quantum percentage of nonreciprocity in the plasmon dispersion to be, 
\bea
\eta^{\cal C} &\equiv& \frac{\Delta\omega_p^{\cal C}(q)}{\omega_p^{0}(q)}\sim\frac{1}{\alpha_{\rm ee}} \frac{W(\beta)}{4\beta}\frac{u}{v_F}\frac{\omega_p^{0}(q)}{|\mu|} ~,\label{Eq_classical}
\\
\eta^{\cal Q} &\equiv& \frac{\Delta\omega_p^{\cal Q}(q)}{\omega_p^{0}(q)}\sim \frac{u}{v_F}\frac{\omega_p^{0}(q)}{4|\mu|}\frac{q}{k_F}~.
\eea
Here, $\omega_p^{0}(q)=|\mu|\sqrt{2\alpha_{\rm ee}}(q/k_F)^{1/2}$ is the long-wavelength plasmon dispersion for graphene without drift flow~\cite{Staubar_2006}. We compare the dependence of both these corrections on different parameters such as wavevector, carrier density, and the interaction strength in Table~\ref{table_I}. We find that the quantum correction to the nonreciprocity is more sensitive to all these parameters. 
\begin{figure*}[t!]
	\includegraphics[width=\linewidth]{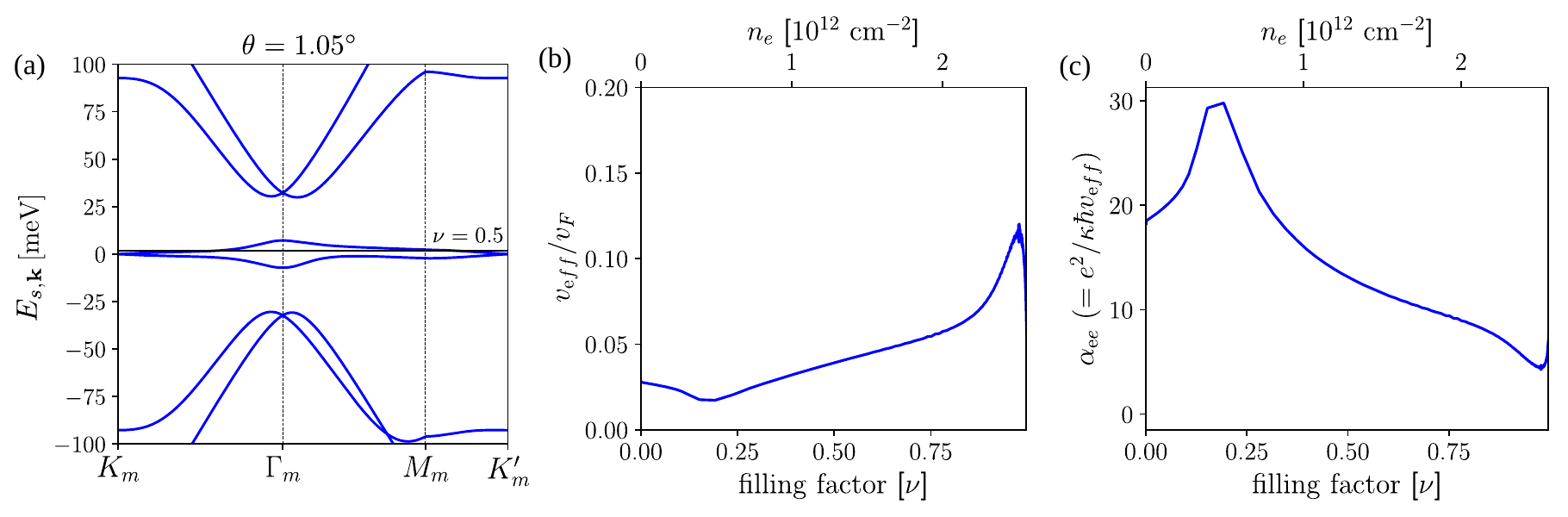}
	\caption{
    (a) Band dispersion of twisted-bilayer graphene near magic angle $\theta=1.05^{\circ}$ for K-valley only. We have used $u_0=79.7$ meV and $u_1=97.5$ meV in the continuum model~\cite{Koshino2018M}. The chemical potential is set at $\mu=1.65$ meV, corresponding to half-filling.
    (b) The variation of effective Fermi-velocity ($v_{\rm eff}$) of TBG with charge filling factor, $\nu$ in terms of Dirac fermion velocity ($v_F\approx 0.87\times{10}^6$ m/s) of single-layer graphene. Here, $n_e$ denotes the corresponding carrier density.
    (c) Variation of the effective fine structure constant, $\alpha_{ee}$ for twisted bilayer graphene with filling factor. $\alpha_{ee} > 10$ for a significant region of the band occupancy. 
    }
 \label{TBG_figure}
\end{figure*}
\begin{table}[t!]
   \centering
   \begin{tabular}{c c c c }
   \hline \hline 
   Graphene & ~$\eta^{\cal C}=\frac{\Delta \omega_p^{\cal C}(q)}{\omega_p^0(q)}$ ~ & ~$\eta^{\cal Q}=\frac{\Delta \omega_p^{\cal Q}(q)}{\omega_p^0(q)}$ ~
  \\
   \hline \hline 
   wavevector  & $q^{1/2}$ & $q^{3/2}$  
   \\
   & & \\
   Carrier density & $n^{-1/4}$  & $n^{-3/4}$
   \\
   & &   \\
   Effective fine structure ($\alpha_{\rm ee}$) & $\alpha_{\rm ee}^{-1/2}$ & $\alpha_{\rm ee}^{1/2}$~
   \\
   \hline \hline 
   \end{tabular}
   \caption{The dependence of classical ($\eta^{\cal C}$) and quantum ($\eta^{\cal Q}$) percentage of plasmonic nonreciprocity for graphene on the wavevector, carrier density, and the effective-fine structure constant. 
   }
   \label{table_I}
\end{table}

In Fig.~\ref{Fig2_SLG}(a), we present the total nonreciprocal plasmon dispersion in graphene for different drift velocities or $\beta$. This includes both classical and quantum corrections. We also show the classical and total percentage of nonreciprocity separately in Fig.~\ref{Fig2_SLG} (b). 
For graphene, the total percentage of nonreciprocity is mainly dictated by the classical contribution with $\eta^{\cal C}$~\cite{DNBasov2021}. The quantum correction is small in the range of experimentally accessible wavevectors. This is because the quantum correction $\Delta \omega_p^{\cal Q}\sim \alpha_{\rm ee} uq^2$ is smaller for small wavevectors ($q< k_F$) with $\alpha_{\rm ee}\approx1$~\cite{Grigorenko2012}. The quantum correction can become significant for a larger wavevector ($q>k_F$), but the plasmon enters the particle-hole continuum region and becomes Landau damped~\cite{Staubar_2006}. This suggests that the quantum nonreciprocity of the plasmon dispersion can become larger in \moire superlattices of graphene, which support more significant $\alpha_{ee}\gg 1$~\cite{Lewandowski2020}, and long-lived plasmon for larger wavevectors. Motivated by this, we investigate drift-induced nonreciprocal plasmon dispersion in the \moire superlattice of twisted-bilayer graphene in the next section.

\section{Large plasmonic nonreciprocity in twisted bilayer graphene}

Moire superlattices, in general, and TBG, in particular, have attracted a lot of attention in near-field optical spectroscopy studies for probing novel collective plasmon modes~\cite{FrankHLKoppen2021, Huang2022, Levitov2019}. Here, we specifically focus on the nature of drift-induced nonreciprocal plasmon in \textit{magic} angle TBG. 

The \moire superlattice has a large periodicity, of the order of tens of nanometers, for a small twist angle ($\theta$). The real space lattice constant is speciifed by $L_{M}=a/[2\sin(\theta/2)]$, where $a\approx 0.246$ nm. This also leads to a smaller \moire Brillouin zone, with a reciprocal lattice vector of magnitude $k_{M}=k_{\rm BG}\sin(\theta/2)$, with $k_{\rm BG}$ denoting the magnitude of the reciprocal lattice for bilayer-graphene (BG). Due to the interlayer electronic coupling and modulation of Dirac fermions by \moire superlattice potentials~\cite{Bistritzer2011, Koshino2018M}, the electronic band dispersion of small angle TBG shows distinct features compared to SLG and Bernal-stacked BG. Near magic angle $\theta\approx 1.05^{\circ}$, the low energy band structure of TBG has four quasi-flat bands (two for valley, and two for spin degeneracy) with minimal bandwidth ($\sim 8$ meV) as shown in Fig.~\ref{TBG_figure}(a). The description of the continuum model Hamiltonian is discussed in Sec.~\textcolor{blue}{S7} of SM~\cite{Note1, Andrei2020, Santos2007G, Atasi_2022, Sinha_B_2022, atasi2022N}. As a consequence, the effective Fermi velocity ($v_{\rm eff}$) of the carriers close to the charge neutrality point becomes around $v_{\rm eff}\sim 0.04v_F$, while SLG has  $v_F=0.86\times{10}^6$ m/s [see Fig.~\ref{TBG_figure}(b)]. As a consequence, the effective fine-structure constant in TBG, $\alpha_{\rm ee}=e^2/(\kappa\hbar v_{\rm eff})$, gets significantly enhanced ($\alpha_{\rm ee}\sim$ $20-30$) compared to SLG having $\alpha_{ee}\approx 1$~\cite{Grigorenko2012}. We highlight this explicitly in Fig.~\ref{TBG_figure}(c), by showing the variation of $\alpha_{\rm ee}$ in the first conduction band with doping. 

\begin{figure*}[t!]
	\includegraphics[width=\linewidth]{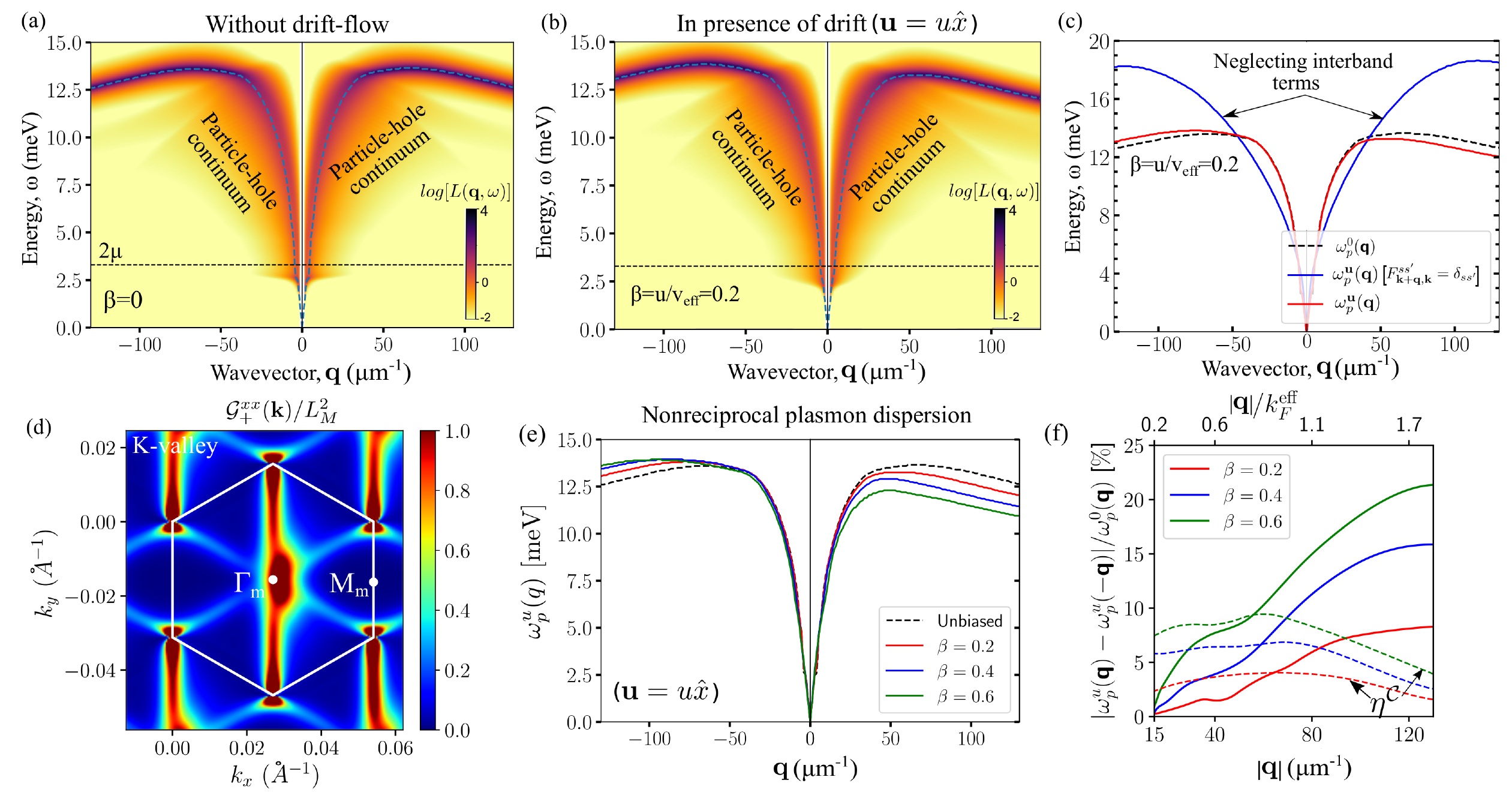}
 	\caption{(a) Colorplot of the loss function spectrum, $L(\bm{{q}},\omega)={\rm Im}\left[-\varepsilon^{-1}(\bm{q},\omega) \right]$ {in log scale} for twisted bilayer graphene (TBG) encapsulated in hBN (with $\kappa= 4.9$~\cite{marco_tbg20}) at half-filling ($\mu=1.65$ meV) 
    without any DC bias. We set wavevector $\bm{q}\parallel {\Gamma_m}$-${\rm{M}_m}$ direction as shown in panel (d).
   The cyan dashed line represents numerically evaluated plasmon dispersion by finding the roots of the dielectric function. (b) Loss function spectrum of biased TBG in the presence carriers drifting with velocity $\bm{u}$ along the x-direction.
   The plasmon dispersion exhibits asymmetry for $+\bm{q}$ and $-\bm{q}$ wavevector. 
   (c) {Nonreciprocal plasmon dispersion (shown by blue line) in TBG, without considering band geometric corrections by setting $F^{ss^{\prime}}_{\bm{k}+\bm{q},\bm{k}}=\delta_{ss^{\prime}}$. This neglects all interband contributions in the density-density response function.}
   (d) Colormap of the quantum metric, ${\cal G}^{xx}_{+}(\bm{k})$ for the first conduction band in the K-valley. (e) Nonreciprocal plasmon dispersion in TBG for different values of $\beta= u/v_{\rm eff} = 0.2$, $0.4$, and $0.6$, respectively.
   (f) Total percentage of nonreciprocity (solid lines) in TBG with different $\beta$. {The horizontal axis at the top of the panel has been scaled in terms of the effective Fermi wavevector $k_F^{\rm eff}\approx 71~{\rm{ \mu m}}^{-1}$.} The dashed lines represent the classical percentage of nonreciprocity ($\eta^{\cal C}$) with band-overlap factor, $F^{ss^{\prime}}_{\bm{k},\bm{k\pm\bm{q}}}=\delta_{ss^{\prime}}$. 
   Here, quantum corrections dominate over classical contributions, particularly in large wavevectors.}
    \label{TBG_figure3}
\end{figure*}

To demonstrate the nonreciprocal plasmons in TBG, we have numerically calculated the loss-function spectrum, $L(\bm{q},\omega)={\rm Im}\left[-\varepsilon^{-1}(\bm{q},\omega) \right]$ including all the intra ($s=s^{\prime}$) and interband ($s\ne s^{\prime}$) transitions for both K and K$^{\prime}$ valley. We present the calculated loss-function spectrum in pristine TBG (without drifting carriers) in Fig.~\ref{TBG_figure3}(a, b). For our calculations, we have used 2D Coulomb potential $V_q=2\pi e^2/(\kappa |\bm{q}|)$, with $\kappa=4.9$ as the background static dielectric constant for hBN~\cite{marco_tbg20}, and $T=5$K. The loss function displays sharp peaks at the plasmon poles and shows the long-lived nature of the intraband plasmons in TBG in the terahertz frequency regime. In contrast to plasmons in SLG, the intraband plasmon mode in TBG lies above the Pauli blocking regions [$\omega_{p}^{0}(q)\gg2\mu$]~\cite{Levitov2019}. {It becomes undamped from particle-hole excitations for large momentum as shown in Fig.~\ref{TBG_figure3}(a).} 
On applying a finite DC bias voltage and enabling drifting charge carriers in TBG, the Fermi surface shifts by momentum $\delta_{\bm{k}}=-\beta k_F^{\rm eff}\bm{\hat{\bm{u}}}$ where $\beta=u/v_{\rm eff}$, and $v_{\rm eff}=k_F^{\rm eff}/m_{\rm eff}$. Here, $v_{\rm eff}$ is the effective Fermi velocity, $m_{\rm eff}$ denotes the effective mass of the carriers, and $k_F^{\rm eff}=\mu/({\hbar v_{\rm eff}})$ is the effective Fermi wavevector for TBG. 
We present the drift current (along x-direction) induced nonreciprocal loss function in the $\bm{q}-\omega$ plane in Fig.~\ref{TBG_figure3}(b). 
The nonreciprocal nature of the plasmon dispersion with $\omega_{p}^{\bm{{u}}}(-\bm{{q}})\ne \omega_{p}^{\bm{{u}}}(\bm{{q}})$ can be clearly seen. 
To investigate the classical correction in the nonreciprocity, we have calculated nonreciprocal dispersion in Fig.~\ref{TBG_figure3}(c) by artificially setting the band-overlap factor, $F^{ss^{\prime}}_{\bm{k},\bm{k}+\bm{q}}=\delta_{ss^{\prime}}$, where $\delta_{ss^{\prime}}$ denotes Kronecker delta function.
{This approximation necessarily neglects all interband overlap terms, as well as band geometric corrections in TBG, mapping the TBG problem to a 2DEG case.} The quantum corrections arise primarily from the quantum metric [see Eq.~\eqref{Eq_A1a}], which dominates at larger wavevectors. {This is because the plasmon mode is undamped in TBG for larger wavevectors, which allows to dominate $\Delta \omega_p^{\cal Q}(\bm{q})$ over $\Delta \omega_p^{\cal C}(\bm{q})$.} We present the distribution of the quantum metric in the 2D Brillouin zone in Fig.~\ref{TBG_figure3}(d). 

Finally, we present the drift velocity dependence of the plasmon dispersion in Fig.~\ref{TBG_figure3}(e) by numerically solving Eq.~\eqref{epsilon_RPA} for different  ($\beta=u/v_{\rm eff}$)  values. We show the corresponding drift velocity dependence of the percentage of the plasmon nonreciprocity in Fig.~\ref{TBG_figure3}(f). {The classical correction overestimates the nonreciprocity for small $q$ values}.
Interestingly, we find a significant increase in the percentage of total nonreciprocity between two oppositely propagating plasmon modes at larger wavevectors, driven predominantly by the quantum corrections [see Fig.~\ref{TBG_figure3}(c)]. 
The quantum Doppler shift-induced plasmonic nonreciprocity can be more than 20\% for $\beta \approx 0.6$. Furthermore, given the low value of $v_{\rm eff}$ in TBG and in other \moire materials in general, achieving a more significant value of $\beta$ in experiments should be feasible~\cite{Alexey2022}. 

Our calculations strongly suggest that TBG and other \moire platforms with relatively flat bands can be good candidates to observe a significant quantum plasmonic Doppler effect. This is enabled by a rather large value of the effective interaction parameter ($\alpha_{ee} \propto 1/v_{\rm eff}$) originating from the smaller band velocities in flat bands. It will be interesting to probe this large nonreciprocity in near-field imaging experiments~\cite {FrankHLKoppen2021, DNBasov2021}. {Further, we note that all our calculations are within the RPA, which misses out on exchange and correlation effect~\cite{giuliani2005quantum}. Generally, RPA works very well for the plasmon dispersion. In fact, the plasmonic Doppler shift, calculated within RPA, explains the experimental dispersion for single-layer graphene reasonably well~\cite{Zhao2021}. However, subtle effects in the plasmon nonreciprocity induced by exchange and correlation effects cannot be ruled out completely.}


\section{Conclusion}
In summary, our investigation into drift-induced nonreciprocal plasmon dispersion in a general quantum system has unveiled an intriguing quantum plasmonic nonreciprocity alongside the well-established classical Doppler shift. The classical correction ($\Delta\omega_p^{\cal C}$) mainly arises from Fermi surface polarization under a DC electric field. In contrast, the quantum correction, ($\Delta\omega_p^{\cal Q}$), stems from the quantum-metric dipole - a fundamental band geometric property of the Bloch wave function.

Explicitly examining single-layer graphene and the moiré superlattice of twisted bilayer graphene, we observed distinct behaviors. In single-layer graphene, the classical term ($\Delta\omega_p^{\cal C}\sim uq$) predominantly governs the plasmonic Doppler shift, with the quantum correction ($\Delta\omega_p^{\cal Q}\sim \alpha_{\rm ee}uq^2$) being relatively smaller owing to its $q^2$ behavior with $\alpha_{\rm ee}\approx 1$~\cite{Grigorenko2012}. Conversely, the quantum metric-induced plasmonic quantum Doppler shift takes precedence in twisted bilayer graphene. This dominance arises from a small band velocity in the flat bands, resulting in a large effective fine structure constant ($\alpha_{\rm ee}\approx 20-30$). Additionally, plasmons in twisted bilayer graphene remain practically undamped even for large values of $q/k_F^{\rm eff}$, allowing the $\alpha_{\rm ee}uq^2$ term in $\Delta\omega_p^{\cal Q}$ to influence the dispersion. Consequently, twisted bilayer graphene and other moiré systems can exhibit a plasmonic nonreciprocity of 20\% or higher for reasonable drift velocities~\cite{Alexey2022}.

This exploration of drift-induced nonreciprocal plasmons in twisted bilayer graphene advances the fundamental understanding of the subject and paves the way for novel optoelectronic applications. Possibilities include the development of plasmonic isolators~\cite{FAN_AIP2009}, one-way waveguides~\cite{Yu_2012}, and optical transmission~\cite{PRL2010Khanikeav}.

\section*{Acknowledgment}
\noindent DD acknowledges the Indian Institute of Technology, Kanpur, for financial support. A.~A acknowledges the Department of Science and Technology for Project No. DST/NM/TUE/QM-6/2019(G)-IIT Kanpur, of the Government of India, for financial support.
We thank Atasi Chakraborty and Debottam Mandal for the valuable discussions. We acknowledge the high-performance computing facility at
IIT Kanpur for computational support. We also acknowledge the National Supercomputing Mission (NSM) for providing computing resources for 'PARAM Sanganak' at IIT Kanpur.
\bibliography{reference}
\end{document}